\documentclass[12pt]{iopart}

\usepackage{graphicx}
\usepackage{dcolumn}
\usepackage{bm}
\begin{document}
\title{Diversity under variability and extreme variability of environments}
\author{Juan G. Diaz Ochoa}
\address{\mbox{Institute for Theoretical Physics, 
University of Bremen, 
Otto Hahn Allee, D-28359 Bremen}}
\ead{diazochoa@itp.uni-bremen.de}
\begin{abstract}
We conducted an investigation of the effect that extreme variability of the individual's environment has in the individual's adaptability and, in general, in the co-evolution of a population. {\it First} we assume that the individuals are a kind of perceptrons with a given memory size that adapt to their environment; {\it second} we consider co-evolution not only of the individual actions but also of the memory size, depending on the individual's fitness; and {\it third} we assume strong variability of the environment. We show that changes in the intensity of the environmental fluctuations introduce a phase transition in the frequency of cooperators and the diversity of the population. We also find out that extreme variability of the environment induces changes in the critical exponents and that this variability can promote more diversity in the population. 

\end{abstract}
\maketitle
\section{Introduction}
An important aspect in ecology is the comprehension of the distribution of populations of different species in space and time as well as the effect of the fluctuating environment in this diversity. Both aspects concern the effort to conserve species, which should consider the extreme heterogeneous distribution of this diversity: from high diversity, like hot spots such as Choco or the West African coast, to less diverse systems, like the German forest. Not only the distribution, but also the extreme changes in the environment have a critical importance in forecasting the effect of extreme phenomena, such as global warming, in ecosystems \cite{Botkin}.

The effect of fluctuating environments has suggested a kind of insurance mechanism in populations. According to this hypothesis, diversity can insure a population against decline in its functioning because there is a large set of actions that can give a response to these fluctuations (See Shigeo Yachi et al. \cite {Yachi}). However, other models show that more diversity cannot resist fluctuations in the environment, because the system fails to find, in a finite time, an equilibrium state (R. May \cite{May}). This dispute is still open and restricted to the distribution of phenotypes. Less attention has been given to the distribution of genotypes among a population. In recent years some theoretical works have investigated the geographic distribution of genetic diversity and, in particular, how some species are unevenly distributed \cite{BarYam}, but this analysis is restricted to single individuals and does not take into account its interplay with the environment. Because there is no clarity of what the effect is of the variability of the environment and individual selectivity on this distribution, it is then necessary to develop adequate models that allow this analysis. In this investigation we will focus on this specific problem.

If we want to consider, from a theoretical point of view, the effect of the environment on the population and, at the same time, the effect of the population on its environment, then a good starting point is to make use of game theoretical tools \cite{Nowak,Moran,Sigmund,Szabo_Fath}. Some models implement a grade of variability to game theory assuming that the individuals can spatially change. Such changes induce effects in the co-evolutive dynamics \cite{Frey}. This approach gives information about the natural behavior that some species may have in its population. For example, several species of birds migrate depending on seasonal circumstances or simply as a natural reaction to concurrence against other individuals.

Other works analyze the effect that the variation of the environment has in a population by means of analytical models considering phenotype switching. It has been shown that stochastic switching can be favored over sensing when the environment changes infrequently \cite{Leibler}. When the environment is considered, its statistics are of central importance and must be taken into account in the population dynamics. Such dynamics have an effect in the analysis of the phenotypic diversity (for social systems see for instance Huberman et al. \cite{Huberman}). But the population's genotypic diversity (or in general the relation to the memory size of individuals) is an overlooked component that has wide-ranging effects in the community structure and ecosystem processes \cite{Crutsinger}. Therefore, it is of fundamental interest to extend the analysis of the population of genotypes (memory sizes) in contrast with the conventional analysis of genotypes.

We focus on two behavioral aspects of the individuals: {\it First}, we consider adaptation and {\it second} we consider the individual ability to select the environment. The first aspect connects the genetics to the adaptability of individuals; in this case, each individual adopts an action as a function of the information it stores from the opponents as well as its own past actions (implying an individual learning process). This characteristic has allowed the construction of models that connect machine learning to genetics \cite{Farmer}. However, this particular aspect can be extended to other applications of game theory where agents with adaptation are modeled \cite{Marsili}. In machine learning the variability of the environment is represented as the noise when the information is processed. Naturally, the adaptability of the individual depends in this case on the individual's memory size. The diversity of the population is, therefore, related to the number of different individuals with different memory size competing in order to increase their utility.


In a theoretical game scenario, the environment could change in a periodical way. Indeed, either the environment can have extreme changes or the individuals can choose the best-suited environment. This last case can have an origin in the change of its spatial position or by changing the interaction with other species. If the individuals can choose their interactions, they must then possess some mechanism that allows them to make choices. In co-evolution a central brain conduces the actions of the individuals attempting to increase the individual's fitness. However, an important part of this argument is the role played by a decentralized brain \cite{McClure}. A choice, which is related to a preference, can be located in two different levels: one (relevant for population dynamics and genetics) is a genetic level; the second (relevant for the behavior of populations of individuals) is a rational level. In the first case it has been found that a large portion of the human genetic code is composed by non-protein-coding introns, which are nucleotide sequences that do not code specific proteins \cite{Hammock}. This is a type of junk DNA called micro-satellites; a key feature is that some aspects in social behavior are apparently governed by repetitive micro-satellites found into the introns. The second level concerns the reaction in the brain within a decision making process, where the integration of emotional states with stored memories influence the way an individual define its choice \cite{McClure}.

The present work is divided in the following parts. In the second part we explain in three subsections the basic definitions of the game: in the first subsection we introduce basic aspects of game theory and individual's selectivity; in the second subsection we explain the mechanism for individual's adaptability; finally in the third subsection we explain the mutation mechanism in the population. In the third part we present the central results obtained within this work. In the final part we discuss these results and we present an outlook of subsequent contributions.

\section{Mathematical Basis and model}

\subsection{Game theory and individual's selectivity}

The actions of each individual and its opponent are rewarded according a pay-off matrix. This pay off represents the fitness of the individual related to a given environment. In several cases, the individual's action can also shape the environment, implying a co-evolution of the population and its environment. Such co-evolution of individual's fitness and individual's action is mathematically represented by means of game theory.  

The action of individual $i$ is represented by an unitary vector ${\bf \sigma^{i}}$, that can be either ${\bf \sigma^{i}}=(1,0)$ for $C$ (Cooperate) or ${\bf \sigma^{i}}=(0,1)$ for $D$ (Defect). The pay-off of individual $i$ relative to individual $j$ is given by the following matrix \cite{Sigmund}
\begin{equation}
U^{ji}={\bf \sigma^{j}}{\bf F_{p}}{\bf \sigma^{i}},
\label{UTIL_I}
\end{equation}
where $\sigma^{j}$ is the action of individual $j$ and ${\bf F_{p}}$ is the game matrix. The pay-off of the agent $j$ is given by $U^{ij}={\bf \sigma^{i}}{\bf F_{p}}{\bf \sigma^{j}}$. In general $U^{ji} \ne U^{ij}$. In the present model only pair interactions are allowed. The fitness matrix is represented by
\begin{equation}
{\bf F} =\left(\begin{array}{cc}
R & Q \\ 
S & P 
\end{array}\right).
\end{equation}
The scenario implemented in this work is for a prisoner's dilemma game, i.e., the game matrix has the values $Q>R>P>S$, where $Q$ is for temptation, $R$ for reward, $P$ for punishment and $S$ for sucker. In this model $Q = 5$, $R = 3$, $P = 2$ and $S=0$. Using eq. (\ref{UTIL_I}), the total outcome for the individual $i$ respect to its opponents can be computed as
\begin{equation}
f^{i} = \sum_{j=1}^{K}U^{ji},
\label{utility}
\end{equation}
and the total outcome of the opponents respect to individual $i$ is given by
\begin{equation}
f'^{i} = \sum_{j=1}^{K}U^{ij};
\label{utility_I}
\end{equation}
where $K$ is the number of neighbors. The utility is a fundamental quantity that should determine the co-evolutionary process. Given that the structure of the pay-off matrix is in general non-commutative, then $f'^{i} \ne f^{i}$. 

\begin{figure}[h!]
\begin{center}
\includegraphics[clip,  width=0.70\textwidth]{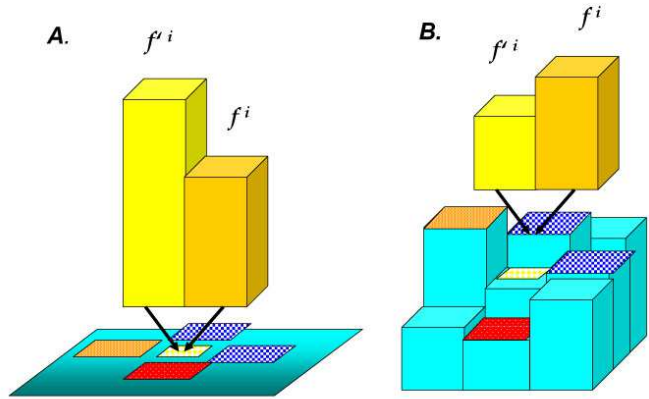}
\caption{Schema of agent's selectivity for individual preferred interactions. In panel {\bf A.} a classic game on a single surface, representing the rigid pay-off matrix, is sketched. In schema {\bf B.} the plastic game is represented. Individuals are allowed to select their individual game, where each square with different relative size represents the actual outcome of the game. In both cases the utility of individual $i$, $f_{i}$, relative to its neighbors, $f'_{i}$, is represented by the two blocks above the diagram.}
\label{Model}
\end{center}
\end{figure}

However, when the individual observes that its game does not increase its fitness, then it can choose either to play a different game or simply not to play (see fig. \ref{Model}). Indeed, the individuals do not interact on a simple surface via a single pay-off matrix, but on a more complex landscape where the pay-off matrix is selected by the gamblers. The definition of non-absolute values in the interactions is made by means of an extended interaction matrix ${\bf F_{p}}$. Furthermore, an additional vector ${\hat \zeta _{k}}$ defines the preferred interaction of the agent $i$. This vector points into each one of the interaction sub-matrices to be fixed in each interaction. The utility matrix for the agent $i$ has in this case the following form \cite{Diaz}
\begin{equation}
U^{ji}=\delta_{l}^{i}{\bf \sigma^{j}}[{\hat \zeta^{l}}{\bf F_{p}}{\hat \zeta^{i}}]{\bf \sigma^{i}}.
\end{equation}
The games are characterized by three kinds of sub-matrices. One sub-matrix is defined under the restriction $2R > T+S$. The second sub-matrix is defined as a chicken game, which means that the mutual defection is the worst possible outcome. Therefore, the values for this sub-matrix are $T'>R'>S'>P'$. The values were arbitrary defined, but in such a way that $T'< R'$, i.e., 'temptation' is not dramatically different to 'reward'. Additionally, $T'<T$, i.e. is a low risk game. A third sub-matrix representing a constant interaction is introduced, with the condition $T''=R''=S''=P''=1$. This last sub-matrix implies for the agents a total risk aversion with low incomes. Hence, the agents interacting in the present model could basically choose between more or less risk. Simultaneously, these three fundamental options allow the game to switch between stable and unstable equilibrium states.

\subsection{Adaptability}
We consider adaptable individuals that select an action according to what they observe from other individuals. The dynamics for the adaptation are represented using methods inspired in machine learning (See for example\cite{Engel,Kirkpatrick}). The individuals store information from the actions of their neighbors as well as their own actions in order to decide a new action.This dynamic implies that the individual is adapting to the whole population. In the present model we consider two layers: $\sigma_{I}$ for the input information into the individual's memory and $\sigma_{O}$ for the individual's output action. In genetics, the first layer contains the information sequence of the grade in which two genes are inherited from a common ancestor \cite{Majewski}.

We define a memory containing the information sequence shared by the individual and its opponent. In order to keep this model as simple as possible we propose a function $\mathcal{S}^{i}(t)$ giving the grade of linkage of the individual's and its opponent's actions (alleles). In a similar way as the genetical affected-sib-pair (ASP) analysis, this function allows us to define a model similar to a one dimensional Ising model \cite{Majewski}, where the probability to obtain an offspring $\sigma_{O}$ depending on the linkage of markers into the individual's memory sequence.

The opponent's as well as the individual's past action at time $t-l$ (offspring at time $t-1$) are defined as an input information (that can be either 1 for $C$ or -1 for $D$, which is related to the allele correspondence to previous offspring). For example the inividual $i$ stores the information of its opponent $j$ in the following way: $\sigma^{j}_{O}(t-1)$ corresponds to the input state $\sigma_{I}^{j1}$, $\sigma^{j}_{O}(t-2)$ corresponds to the input state is $\sigma_{I}^{j2}$; in general $\sigma^{j}_{O}(t-l)$ corresponds to the input state is $\sigma_{I}^{jl}$. The individual $i$ can store into its memory $M_{i}$ actions ($\sigma_{I}^{jM_{i}}$) of its opponent $j$ and $M_{i}$ own actions ($\sigma_{I}^{iM_{i}}$), where $M_{i}$ is the memory size of individual $i$. The function $\mathcal{S}^{i}(t)$ is defined as

\begin{equation}
\mathcal{S}^{i}(t)= M_{i}\sum_{l=1}^{M_{i}} \sigma_{I}^{jl}\sigma_{I}^{il},
\end{equation}

where the strength of the connectivity -between the individual's and its opponent's markers- is directly proportional to the memory size of the individual. Clearly, this is an oversimplification. However, we want to understand the effect that the fitness of the individual has on its adaptability; given that the memory size depends on this fitness through a game (see next subsection), then this option is a simple way to make an analysis of the interaction between adaptability and fitness.

\begin{figure}[h!]
\begin{center}
\includegraphics[clip,  width=0.65\textwidth]{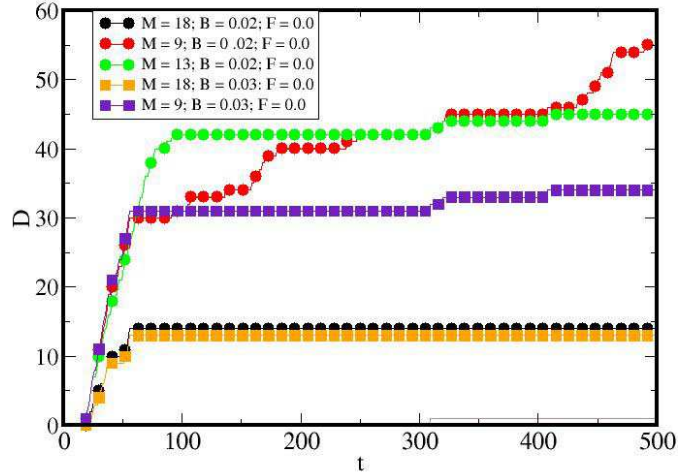}
\caption{Defect frequency of a pair of individuals with different memory size as a function of time for different learning parameters $\beta = 1/\Lambda$. These curves represent tha adaptation process of individuals with different memory sizes. The larger the memory, the larger the probability that an individual remember past actions from its neighbors. Hence, larger memories adapt in few time steps to a cooperate behavior. For $\Lambda \rightarrow \infty$ the relaxation time increases.}
\label{fig1}
\end{center}
\end{figure}

The second layer is the output of the information process: the individual's action, i.e. the individual's offspring, that can be either ${\bf \sigma_{O}^{i}}=(1,0)$ for $C$ (Cooperate) or ${\bf \sigma_{O}^{i}}=(0,1)$ for $D$ (Defect). The transition probability is given by
\begin{equation}
W^{i}_{t \rightarrow t+1}=\frac{e^{-\beta \Delta \mathcal{S}^{i}}}{T},
\end{equation}
where $\beta=\frac{1}{\Lambda}$. $\Lambda$ represents the extend of genetic linkages between markers and the effect of random genetic and environmental efects has on allele sharing \cite{Majewski}, $T$ is the period of time and $\Delta \mathcal{S}^{i} = \mathcal{S}^{i}(t) - \mathcal{S}^{i}(t-1)$.

The variability of the environment is related to the level of noise of the environment (similar to the case of a teacher in a noisy classroom \cite{Engel}). For $\Lambda \rightarrow 0$ the fluctuations of the information are small and the individual's actions relax to a cooperative behaviour; a case that can be related to a frozen system. Otherwise, if $\Lambda \rightarrow \infty$, then the fluctuations of the system increase, avoiding the individual to find in a short time a stable answer. The relaxation time depends on the adaptation parameter $\beta$ and the memory size $M$. The larger the memory, the more stable is the answer and the shorter is the relaxation time. This case is shown in Fig.\ref{fig1} for $M=18$ vs. $M=13$ and $M=9$ for three different parameters $\beta = 0.02$ and $\beta = 0.03$.

\subsection{Mutation of memory size}
The evolutionary dynamics is represented by a mutation process of the individuals. Each individual in the lattice is characterized by its memory $M^{i}$. If the individual's fitness is not large enough, its memory size can mutate into a new memory size $M'^{i}$. In this mutation process we assume that the individual has no lost of information; the additional bit of information added to the memory after the mutation process is assigned in a random way.

This mutation process has the following form \cite{Moran, Nowak_I}: (i) {\it selection} -an individual is selected for reproduction with a probability related to its fitness-; (ii) {\it reproduction} -the individual produces one offspring-; (iii) the offspring replaces a randomly selected individual. The transition probability for the evolution of the memory size $M^{i}$ has the following form:
\begin{equation}
V_{M \rightarrow M'}^{i}(t)=\Theta(f^{i}(t) - f'^{i}(t))\chi^{i}(\nu),
\end{equation} 
where $\Theta(x)$ is a Heaviside step function, $f^{i}$ is the outcome given by eq. (\ref{utility}) and $f'^{i}$ is the outcome of the opponents to the individual $i$, given by  eq. (\ref{utility_I}). The function $\chi^{i}(\nu)$ is the probability that the memory $m^{i}$ changes, where $\nu$ is the mutation frequency.
  
\section{Results}

We want to restrict the analysis to the effects that fluctuations of information have in the population and for this reason we avoid an analysis of a population in a complex topology. The population is therefore represented in a square lattice with periodic boundary conditions, where individuals interact with four individuals selected from their eight nearest neighbors. Along this work the results corresponds to computations performed in lattices with $60 \times 60$ individuals. We assume a co-evolutive process that starts from an initial configuration of individuals with small memories.

\begin{figure}[h!]
\begin{center}
\includegraphics[clip,  width=0.90\textwidth]{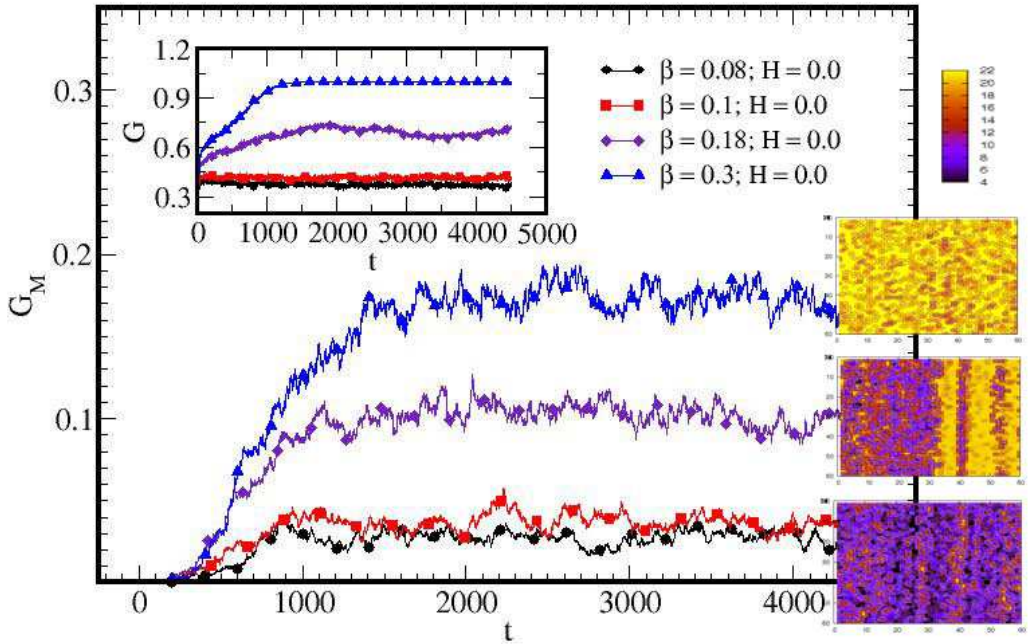}
\caption{Plot of the time dependence of average memories as a function of the time for different learning parameters $\beta$. The snapshots of the different final configurations are also shown. In the inset of the figure, the frequency of cooperators is plotted.}
\label{fig2}
\end{center}
\end{figure}

First we make an analysis of the relaxation of the average number of individuals with a given memory size and of the relaxation of the frequency of cooperators. The density of individuals with a given memory size $\rho^{m}$ is given by
\begin{equation}
\rho^{m}(t) = \frac{\sum_{i=1}^{N}m^{i}(t)}{N},
\end{equation}
where $N$ is the total number of individuals in the population. Using this quantity, the mean of the memory size in the system $G_{M}$ is defined as the average of $\rho^{m}$ and is given by
\begin{equation} 
G_{M}(t)=\frac{\sum_{m} \rho^{m}(t)}{M}.
\end{equation} 
where $M$ is the maximal memory size observed in the system. The frequency of cooperators is also defined by the probability density of individuals adopting the action $C$ along the game,
\begin{equation}
\rho^{i}=\frac{\sum_{c}\rho^{i}_{c}e^{-\beta \mathcal{H}^{i}}}{Z^{i}},
\label{DENSITY_AC}
\end{equation} 
where $Z_{0}$ is the partition function on the storage distribution of the individual $i$ given by
\begin{equation}
Z^{i}=\sum_{c}e^{-\beta \mathcal{H}^{i}}.
\end{equation}
This partition function is valid into the phase space defined by the individual's memory. The distribution function of elements of class $C$, $G$, in the whole population is given by
\begin{equation}
G(t) = \frac {\sum_{i=1}^{N} \rho^{i}(t)}{N},
\end{equation} 

The analysis of the results must account for the distribution of phenotypes (distribution of actions) and the distribution of memory sizes. Both distributions must be analyzed as a function of $\beta$ and as a function of the variability of the environment (in this work this variability is defined as a {\it plasticity} of the game). It is important to remark that the game has direct influence on the memory size of the individuals, not on their actions. The actions are indirectly affected by the individual's memory and directly affected by the dynamics of the memory, which reach an equilibrium state after infinite simulation steps according to the dynamics defined for the individual's actions (Fig. \ref{fig2}). 

First we want to make a characterization of the system. In Fig. \ref{fig2} we present the distribution function of the mean memory $G_{M}$ as a function of the time for different $\beta$ parameters. The higher the value of $\beta$, the more complex the distribution of memories in the population. Given that the simulation starts from some initial distribution of small memories, the system requires some relaxation time until the system reaches an equilibrium state. The relaxation time is also a function of $\beta$.

The parameter $\beta$ is related to the level of noise that avoid the agents adapting to their neighborhood. Small memories adapt much slower than large ones. Hence, a change in the phase behavior of the number of cooperators (or memory size) depends on the noise level that the system is exposed to, i.e. large memories have a better chance to survive if the noise level is lower. 

The noise has important implications in the evolution of a population: more adaptable individuals are best suited to environments with low levels of noise. In the same figure \ref{fig2} we present an analysis of the frequency of cooperators $G$ as a function of the time for different parameters $\beta$. Given that more adaptable individuals tend to cooperate, then the frequency of cooperators increase when $\beta$ also increases, i.e. a population with more cooperators can survive in an environment with less noise. 

\begin{figure}[h!]
\begin{center}
\includegraphics[clip,width=0.70\textwidth]{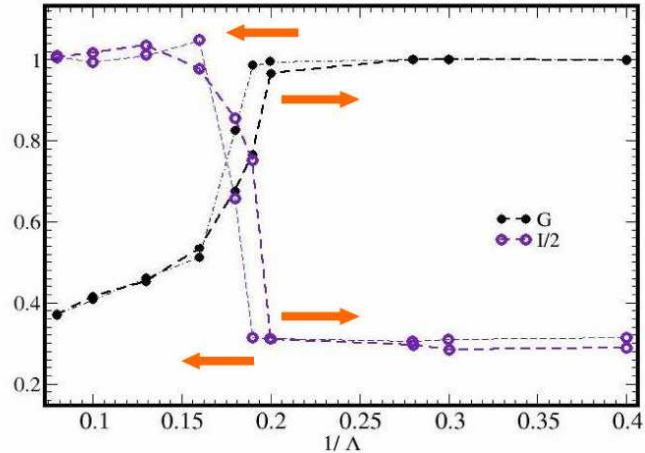}
\caption{Diversity $I$ and frequency of cooperators $G$ as a function of the inverse of the variability of the system $\Lambda$.}
\label{Mem_all}
\end{center}
\end{figure}

Finally three snapshots of the distribution of the memories in the lattice at large times are presented. For $\beta$ above 0.18 the population is dominated by large memories. Below this value we observe dominance of individuals with small memory sizes. The interesting aspect in this distributions is that at $\beta = 0.18$ there is a separation between large and small memories. This kind of phase separation indicates that the system has a behavior similar to a phase separation of first order. 

In the figure Fig. \ref{Mem_all} we present the frequency of cooperators $G$ and the diversity of the population $I$ as a function of the noise level $\beta = 1/\lambda$, where the diversity were defined as
\begin{equation}
I = -\sum_{m=1}^{M} \rho^{m} \ln \rho^{m},
\label{Shannon}
\end{equation} 
where $M$ is the maximal memory size that can be defined into the system. There are several measures for diversity \cite{Simpson,Kleidon,Krebs}. We adopt a Shannon's index, which provides information not only about the population richness, but also the composition of the community \cite{Kleidon, Krebs}.

When the parameter $\beta$ is increased and thereafter is decreased we observe a hysteresis effect. In Fig. \ref{Mem_all} the simulation is started for systems consisting on individuals with relatively small memories (arrow pointing to right). Each point was computed for the same initial configuration and different temperatures. A second group of simulations starting from initial configurations of agents with large memories were also started (arrow pointing to left). In this figure the frequency of co-operators and the re-scaled diversity index $I $ are shown.  

\begin{figure}[h!]
\begin{center}
\includegraphics[clip,  width=0.70\textwidth]{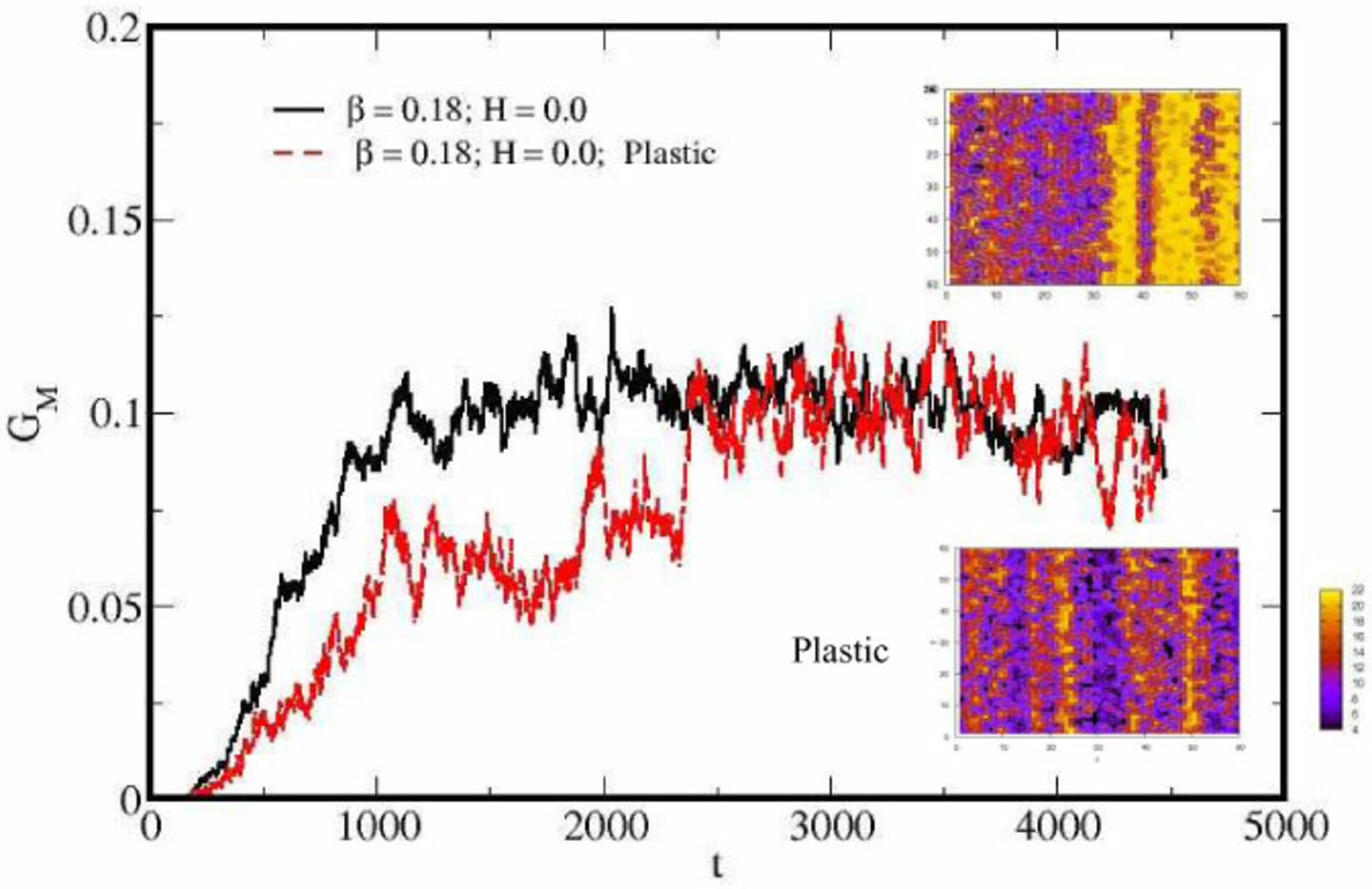}
\includegraphics[clip,  width=0.70\textwidth]{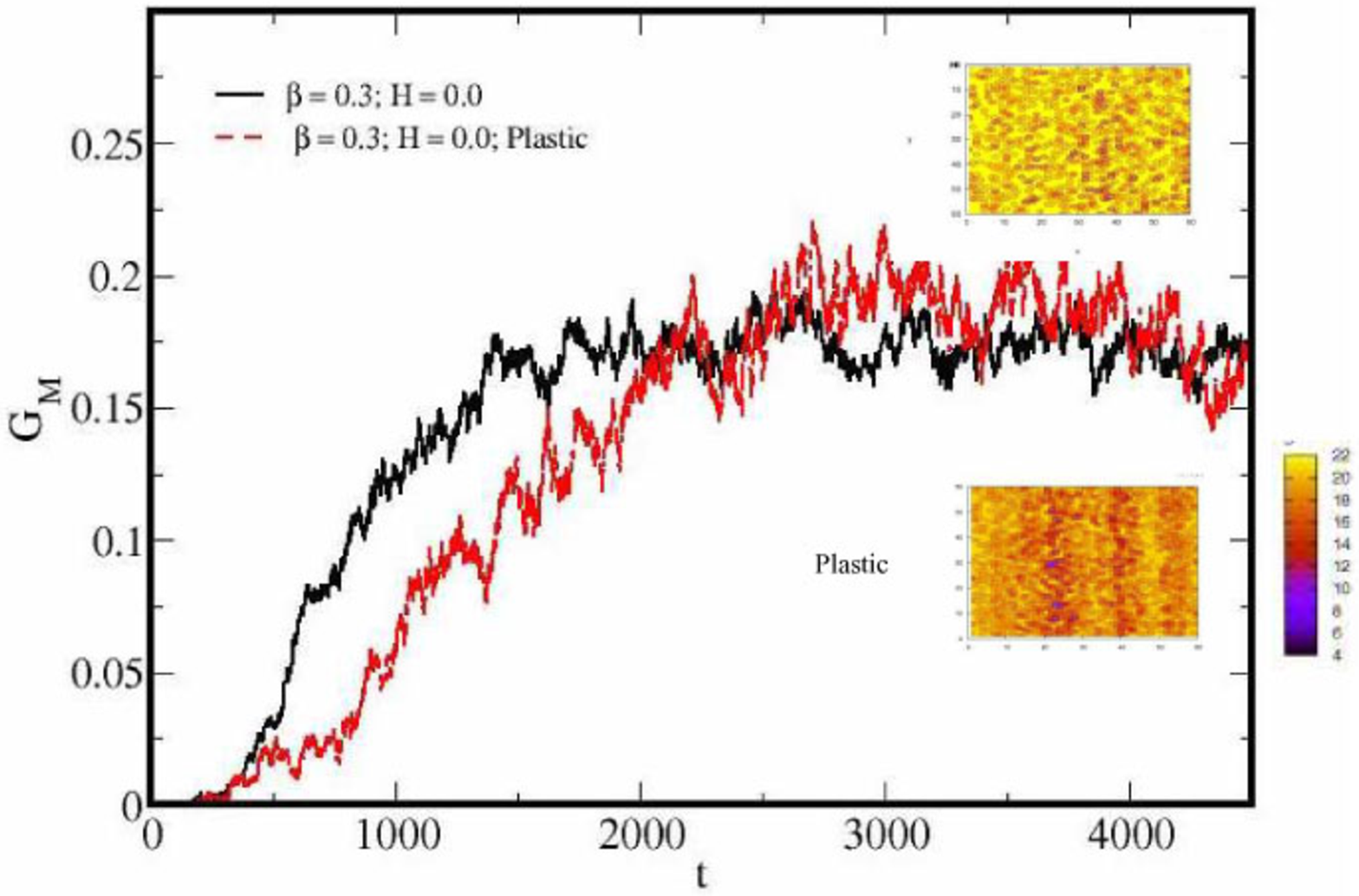}
\caption{Average number of memories $G_{M}$ as a function of time. The first plot corresponds to $\beta = 0.18$ and the second to $\beta = 0.3$. In both cases plastic and non-plastic interactions are plotted. In the insets the corresponding snapshots for the final configurations are also shown.}
\label{fig3}
\end{center}
\end{figure}

The first aspect of this plot is the high diversity contrasted to the low frequency of co-operators for $1/\Lambda < 2$. When $1/\Lambda \sim 2$ both curves show a singularity. Above this critical value the value of diversity is now larger, whereas the frequency of co-operators decreases. The second aspect is the non-reversibility of the curves. This could be related to a phase transition of first order (formation of phase separation between individuals of different memory size) and to memory effects related in a direct form to the model of perceptons for the individuals.  

In order to make an analysis of the plasticity of the interactions in the population we make a comparison of the mean memory as a function of the time for fixed $\beta$ parameters. The remarkable result is the similitude between plastic and non-plastic games. However, the examination of the distribution of memories in the snapshots in the figure \ref{fig3} shows that the separation between low and large memories dissolves.

This separation can be an artifact. The simulation starts in this case from an initial configuration with low memories; the aggregation of large memories has, however, relation to the willingness to cooperate of the individuals. When the environment is extremely variable, large memories have a worse chance than small memories. In such a case, mixed and less structured distributions emerges

It is indeed possible to suggest some effects of the plasticity of the interactions in the relaxation process at short to intermediary time regimes. This result is also confirmed in Fig. \ref{fig3}. In this case the relaxation of the distribution function of the mean memory has been computed as a function of $\beta$ for plastic as well as non-plastic pay-off matrices. The principal conclusion is that the plasticity introduces a retardation in the relaxation of phenotypes into the population. 

In particular for $\beta = 0.18$ is possible to observe further relaxation for $t > 2000$. For this parameter, close to $\Lambda_{c}$, the system with plastic interactions requires more time to overcome fitness barriers before the system relaxes, i.e. this is a metastable state. In contrast to this result, the system with rigid interactions has a simple relaxation curve. For $\beta = 0.3$ the system with plastic interactions reaches an equilibrium in the mean number of memories along a simple relaxation curve.

\begin{figure}[h!]
\begin{center}
\includegraphics[clip,width=0.60\textwidth]{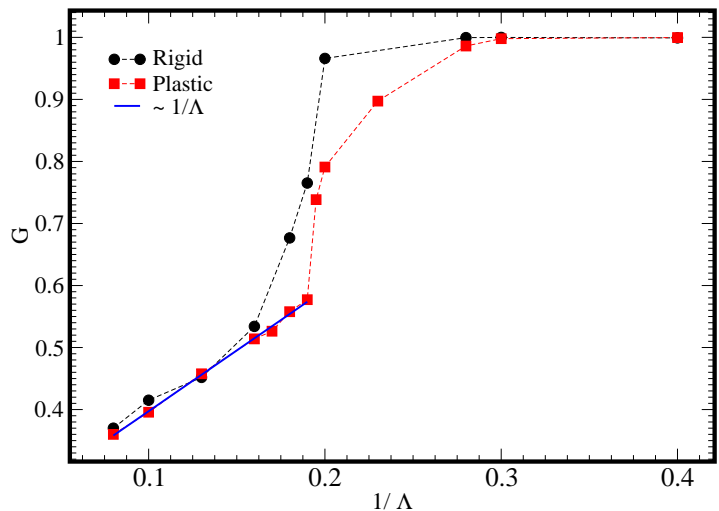}
\caption{Frequency of cooperators as a function of the learning parameter $\beta = 1/\Lambda$ for plastic and non-plastic interactions. The dashed lines are guides for the eyes.
}
\label{fig4}
\end{center}
\end{figure}
We want to make a close analysis of the difference that flexible interactions may introduce into the diversity and frequency of cooperators. In Fig \ref{fig4} the frequency of cooperators $G$ as a function of $1/\Lambda$ is presented. Both kind of games preserve the critical point and confirm that it does not depend on the plasticity of interactions inside the game. However, the phase behavior for rigid games is different to plastic games. In the last case the frequency of cooperators is proportional to $(1/\Lambda)^{\gamma}$, $\gamma = 1$, below the critical point, where there is a singularity. Above the critical point the phase behavior has a polynomial form. In contrast to this behavior, the rigid game has a kind of polynomial behavior below the critical point, with a critical exponent $\gamma$ larger than one; above $\Lambda_{c}$ the frequency of cooperators reach a plateau. 

\begin{figure}[h!]
\begin{center}
\includegraphics[clip,width=0.75\textwidth]{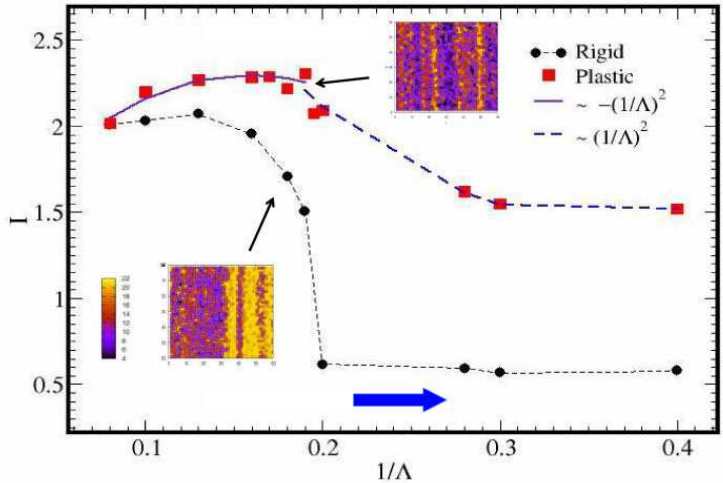}
\caption{Diversity index as a function of the learning parameter. The snapshots of two configurations for plastic and rigid interactions just at the critical point is also included. The blue arrow below the plot represents the direction where $1/\Lambda$ increases.}
\label{fig5}
\end{center}
\end{figure}

The diversity also has a different phase behavior for plastic games (See Fig. \ref{fig5}). For rigid games and $1/\Lambda < 1/\Lambda_{c}$ the phase diagram of the diversity has a polynomial behavior. At the critical point there is a singularity and the diversity reaches a plateau. When we consider plastic games we observe again a remarkable difference with rigid games, in particular the value of the diversity is larger. This result confirms that the variability in the pay-off matrices can increase the diversity of the distribution of memories in the population. The second characteristic is that at $\Lambda_{c}$ there is a cross over in the phase diagram of $I$. Above this value there is a smooth relaxation of the value of the diversity, until it reaches a plateau. The snapshot of the distribution of memories at the critical point for both dynamics can explain the different behavior of both phase diagrams: whereas in the rigid case a phase separation between memories of different sizes emerges, in the plastic case there is a kind of solution of large and small memories, i.e., the plasticity of the interactions destroy the phase separation between memory sizes.

\begin{figure}[h!]
\begin{center}
\includegraphics[clip,width=0.60\textwidth]{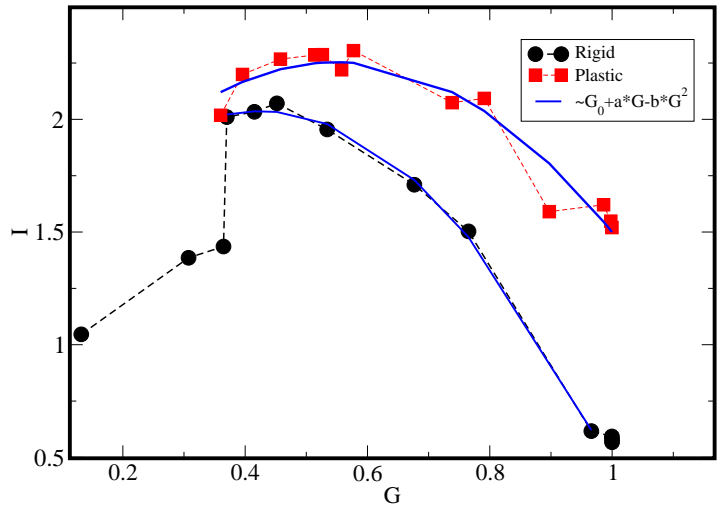}
\caption{Diversity as a function of the frequency of cooperators in the system. The line in black circles corresponds to single rigid interactions. The line in red squares corresponds to plastic interactions. The dashed lines are guides for the eyes.}
\label{Div_coop}
\end{center}
\end{figure}

We establish a relation between the diversity index and the frequency of co-operators. The black line in circles in figure \ref{Div_coop} represents a system with rigid interactions. The red line in squares represents plastic interactions. When the interactions are rigid it is possible to establish a relation that relates in an inverse proportion the diversity $I$ to the frequency of co-operators $G$. This result is as a first approach a natural characteristic of a system with individuals doing an imitation of the actions of their neighbors. According to the rules of the system, cooperative individuals require larger memories. Hence, if the system is more cooperative, then it is also more homogeneous because it is dominated by individuals with larger memory sizes, implying at the same time less diversity. When the diversity $I$ increases, then the willingness to cooperate decreases and a fusion of co-operators and defectors takes place.

The frequency of defectors increases when $G \rightarrow 0$ and simultaneously the diversity index $I$ decreases. Therefore, the phase diagram shows a maximal value of $I$ when the frequency of co-operators and defectors is equilibrated. Below some particular value of $G$ the diversity of the system again decreases. This is because defectors dominate the system.

The implementation of flexible utilities introduces a dramatic change in the behavior of the system. If we allow the individuals to define their best outcome, then the diversity increases without changing the frequency of co-operators $G$. In particular above $G \sim 0.4$ the phase diagram can be fitted using a function of the form $d = G_{0}+a \times G - b \times G^{2}$. We observe that $b_{plastic}\rightarrow 0$, i.e., the contribution of the quadratic term of this function decreases when the game is plastic. Assuming an ecosystem composed of different kinds of individuals, with different internal characteristics, this result proposes a mechanism that shifts the original phase diagram obtained for rigid interactions, helping the system to increase its frequency of diversity without bringing about a change in the frequency of cooperativity.

This result could bring some clues in the comprehension of the diversity of ecosystems. Assume for instance an ecosystem with rigid utilities together with an ecosystem with flexible interactions, both separated by a thin barrier. Both systems have the same intensity of fluctuation $\Lambda$. However, one system shows more diversity than the other one, despite both being very close. In this example the internal characteristics of the environment (the flexible utilities that can be selected by the individuals versus rigid utilities), and not the simple variability of the environment, are fundamental for the estimation of the value of diversity.

\section{Discussion}
The main result is the effect that variability of the pay-off matrix has in the co-evolution of large memories. In this work, as well as in previous works, we have shown that the larger this variability, the lower the chance that large memories have to co-evolve. If the external information source promotes the evolution of small memories, then this effect is more dramatic. 

This result could be surprising but it is, however, very plausible. The larger the memory the more information an individual can store from previous events. This process requires a very stable environment. Once the environment fluctuates, the memory does not represent any advantage in evolution, producing an increase of the fitness of small memories. Indeed, this observation depends on the correlation of the fluctuations of the environment, i.e. for correlated fluctuations the memory represents again a clear advantage \cite{Leibler}. If moreover the external information favors small memories, then the diversity is again out of balance: instead of the evolution of large memories, the dynamics of the population show a prevalence of small memories. Hence, variability as well as information are two important parameters that affect the development of a population.

The plasticity in the interactions also has profound implications in the phase behavior of the system. First, it introduces additional fitness barriers that the system must overcome until it relaxes. Moreover, the plasticity deteriorates the phase separation between large and small memories, generating a much better mixture of individuals. From a qualitative point of view this also is an improvement of the diversity, because it helps to increase the entropy of the system. We can also establish a dependence of the diversity as a function of the number of cooperators. We can determine in this phase diagram a critical frequency of cooperators where the diversity is maximal; we also can observe an increase of diversity when the system is plastic. However, for a critical value of frequency of cooperators the phase behavior for plastic games converges to the phase behavior for rigid games, i.e. extreme variability not necessary improves the diversity of the system.

The results obtained in this investigation could be used as a mechanism that can improve diversity in a population. The implications range from ecology to social science and could explain why genetic diversity can be found in certain systems. In the frame of this investigation, the answer to this question is that diversity could be larger when individuals are allowed to decide their game, i.e., when individuals can search for their best interaction with other individuals in an ecology. 

In social sciences it could be an interesting result for the analysis of cooperation. Some investigations have shown, for instance, that more cooperation requires less diversity \cite{Putnam}. However, the result presented in Fig. \ref{Div_coop} shows that more cooperation can allow more diversity if agents can define their pay-off. Assuming adaptable individuals, this result implies that diversity and cooperativity depend on the degree of freedom that individuals have to move in a fitness landscape.

\ack I want to thank Hauke Reuter for very useful remarks.
\bigskip


\end{document}